\documentclass[pra,twocolumn,showpacs,letterpaper,showpacs,superscriptaddress]{revtex4}
\usepackage[dvips]{graphicx}
\usepackage{amsmath,amssymb,amsfonts,latexsym,color,dcolumn,bm}
\usepackage{verbatim}

\begin{document}
\newcommand{\beq}{\begin{equation}}
\newcommand{\eeq}{  \end{equation}}
\newcommand{\bea}{\begin{eqnarray}}
\newcommand{\eea}{  \end{eqnarray}}
\newcommand{\bit}{\begin{itemize}}
\newcommand{\eit}{  \end{itemize}}
\newcommand{\jmax}{j_{\text{max}}}

\providecommand{\abs}[1]{\left\lvert#1\right\rvert}
\providecommand{\norm}[1]{\lVert #1 \rVert}
\providecommand{\moy}[1]{\langle #1 \rangle}
\providecommand{\bra}[1]{\langle #1 \rvert}
\providecommand{\ket}[1]{\lvert #1 \rangle}
\providecommand{\braket}[2]{\langle #1 \rvert #2 \rangle}

\title{Molecular orientation entanglement and temporal Bell-type inequalities} 

\author{P. Milman} 
\affiliation{Laboratoire de Photophysique Mol\'{e}culaire du CNRS, Univ. Paris-Sud 11, 
B\^{a}timent 210--Campus d'Orsay, 91405 Orsay Cedex, France}
\author{A. Keller} 
\affiliation{Laboratoire de Photophysique Mol\'{e}culaire du CNRS, Univ. Paris-Sud 11, 
B\^{a}timent 210--Campus d'Orsay, 91405 Orsay Cedex, France}
\author{E. Charron} 
\affiliation{Laboratoire de Photophysique Mol\'{e}culaire du CNRS, Univ. Paris-Sud 11, 
B\^{a}timent 210--Campus d'Orsay, 91405 Orsay Cedex, France}
\author{O. Atabek} 
\affiliation{Laboratoire de Photophysique Mol\'{e}culaire du CNRS, Univ. Paris-Sud 11, 
B\^{a}timent 210--Campus d'Orsay, 91405 Orsay Cedex, France}


\begin{abstract} 
We detail and extend the results of [Milman {\it et al. }, Phys. Rev. Lett. {\bf 99}, 130405 (2007)] on Bell-type inequalities based on correlations between measurements of continuous observables performed on trapped molecular systems. We show that, in general, when an observable has a continuous spectrum which is  bounded, one is able to construct non-locality tests sharing common properties with those for two-level systems. The specific observable studied here is molecular spatial orientation, and it can be experimentally measured for single molecules, as required in our protocol.  We also provide some useful general properties of the derived inequalities  and study their robustness to noise. Finally, we  detail possible experimental scenarii and analyse the role played by different experimental parameters. 

\end{abstract} 


\maketitle
\section{Introduction}
Quantum mechanics allows for the existence of states without any classical correspondence. Examples of such states are entangled states, that may appear when describing the total state of a many particle  system, or when describing different degrees of freedom of a single particle. Entangled states cause debate because they can present properties which contradict our classical intuition. In addition,  some of these properties can be use to increase the security of quantum communication and the efficiency of  algorithmic protocols when compared with  classical techniques. This is why so much attention has been payed to problems belonging to the foundations of quantum mechanics, and in particular to quantum entanglement, which is believed to be one of its main traits \cite{SCHRO}. One of the most important discussions on entanglement and its conflicts with classical physics concerns the realism and (non-)locality of quantum physics. As pointed out by Einstein, Podolski and Rosen in the so called EPR \cite{EPR} paradox, entangled states are closely connected to apparent contradictions between quantum mechanics and fundamental physical assumptions. Even if experimental evidence has been obtained to support quantum mechanics against the EPR criticism \cite{ASPECT}, it is still a matter of debate if such experiments were realised in the ideal conditions, closing all the loopholes, so as a definitive conclusion can be reached. At the same time, from the fundamental point of view, identifying precisely which essential quantum properties are involved in local realism violation is still an open problem. This is why extending local realism tests to different physical systems and different physical scenarii still presents so much interest.

A common property of non-local states is entanglement.  In spite of its importance and  consequences in different physical contexts, it remains an open question to determine whether a general quantum system is entangled or not and to quantify the degree if entanglement of a given state. The problem has been solved for some particular cases, as for a general bipartite system of dimension  ${\cal H}_{2}\otimes {\cal H}_{2}$ and ${\cal H}_{2}\otimes {\cal H}_{3}$ \cite{PERES}, where ${\cal H}_d$ is the one particle Hilbert space of dimension $d$. In such cases,  necessary and sufficient conditions for telling whether a given state is entangled or not exists. In other particular cases, or subspaces,   one can also find necessary and sufficient conditions. For instance,  entanglement in bipartite pure states can always be recognised and quantified irrespectively of each parties' dimension. However, when  dealing  with arbitrary states, including the more realistic mixed ones, only necessary conditions for separability (non-entanglement) can be provided. A notion that will be useful in the following of this paper is the one of {\it entanglement witnesses}, defined as an operator $\hat W$ for which the expectation value $\langle \hat W \rangle \leq S$ for all separable states. This ensures that the state is entangled if  $\langle \hat W \rangle > S$. On the contrary, the case $\langle \hat W \rangle \leq S$ \cite{WITNESS} is not conclusive. Examples of entanglement witnesses that are also useful for fundamental tests of quantum mechanics are Bell-type inequalities \cite{WITBELL}, which are the main scope of this paper.

Bell-type inequalities are composed of combinations of observables that, when measured, allow for setting a board between crucial aspects of quantum theory and the classical one. They were formulated by J. S. Bell \cite{BELL} as a late reply to the EPR criticism to quantum physics \cite{EPR}. Bell inequalities aim at answering the question: is quantum mechanics a local and realistic theory?  In order to do so, they combine correlations between local measurements realised in a multiparty system. Since their first formulation, several other inequalities have been proposed, some studying the same type of problem as Bell, other generalising locality and realism tests to many observers and many possible observables. There are a number of Bell type inequalities, and classifying all of them is a work by itself \cite{GISIN2}.

 The inequalities derived here follow the original formulation of Clauser, Horne, Shimony and Holt (CHSH)  \cite{CHSH}, and deal with the following scenario: two observers $A$ and $B$ perform local measurements on a bipartite system. Each observer can chose among two experimental set-ups ($a$ and $a'$ for $A$ and $b$ and $b'$ for $B$). Each measurement performed by $A$ and $B$ can give only two outcomes. At its origin, CHSH inequalities have been formulated for a pair of spin $1/2$ particles or equivalent two--level systems. In the framework of local hidden variable (LHV) theories the measurement outcomes correlation statistics which must fulfil:
\begin{equation}\label{chsh}
|\langle \sigma_a \sigma_b\rangle+\langle \sigma_a \sigma_{b'}\rangle+\langle \sigma_{a'} \sigma_b\rangle-\langle \sigma_{a'} \sigma_{b'}\rangle|\leq 2,
\end{equation}
where $\sigma_{\alpha}$ is the Pauli matrix in the $\alpha$
direction.  Briefly, in a LHV theory,  one  assumes that measurements performed by each observer are independent and their outcomes have a probability distribution which is a product of independent probabilities for each subsystem. Such probabilities can also depend on some random local variable. Details are discussed in many works, as \cite{GISIN}, for instance. It can be shown that some entangled states can  violate (\ref{chsh}), and this experimental violation was observed using photon pairs entangled in polarization~\cite{ASPECT}. In this case, directions $a$, $a'$, $b$ and $b'$ refer to different 
orientations of polarizers placed before the detectors, determining the direction of the Pauli matrix that is measured. It was shown by Cirelson \cite{CIRELSON} that the maximum value of (\ref{chsh}) is $2\sqrt{2}$, and it is easy to verify that this maximal violation can be obtained with maximally entangled states. 

Inequalities as (\ref{chsh}) have proved to hold for two-level systems or equivalent ones. By ``equivalent ones", we also include Bell type inequalities involving  a continuum of possible measurement outcomes   that are dichotomised and transformed into a two outcomes measurement set-up. Dichotomization works as follows: one splits in two classes the range of possible measurement  outcomes. All results obtained lying in one of the  classes is identified to a given value ($+$ or $-$) and the results obtained in the complementary space are associated to the opposite sign. Some examples of systems where this can be sucesfully done are optical fields \cite{BUZEK,EPJD}.  An interesting problem is to derive Bell-type inequalities for continuous variables without appealing to dichotomisation.  Cavalcanti {\it et al.} found a way out by using second moment correlations instead of first moment ones, as done in CHSH-type inequalities as (1) \cite{REID}. Here, we deal with this problem in a different way: by using bounded observables, one can still use CHSH-type inequalities to detect non-local properties.  In this case, a Cirelson bound depending on the norm of the measured observable can also be derived, even if, at least for the specific case treated in this paper, we have not shown yet that it can be attained. An interesting property of the inequality discussed in this work is that it can be used not only for infinite  dimensional systems, but also in $N$ levels systems, where $N$ is a finite number. In this case, the numerical value of the bound  splitting between a local theory and non-local one depends on the maximal eigenvalue of the measured observable.

 Up to now, several studies have been made on Bell-type inequalities in different contexts. The general conclusion is that the subject still presents several open questions and no general theory is available. In particular, a number of intriguing features coming out from such studies somewhat contradict our acquired quantum ``intuition":  in  \cite{ZUKOWSKI} it is  numerically shown that bipartite multidimensional states may violate locality tests more than two qubits. For two qubits, it has been proven that  a maximal  violation exists, and it is  given by the Cirelson bound\cite{CIRELSON}, as will be discussed hereafter. Acin, Gill and Gisin showed, some years later, also using numerical tools, that the maximal violation for pure bipartite multidimensional systems is not obtained for maximally entangled sates \cite{ANTONIO}. General rules relating local realism violation and entanglement have not yet been found, even if it has been proven that all non-local states are entangled in some way. This fact can be easily understood with the help of entanglement witnesses. It has been shown in \cite{WITBELL} that Bell-type inequalities are entanglement witnesses, and all non-local states are entangled.  

The inequalities studied in this paper are based on molecular  spatial orientation correlations measurements \cite{PRL} instead of spin-like observables. However, the same type of idea can be generalised to other continuous bounded observable. We have shown that they can be implemented using time delayed measurements of correlations between the spatial orientations of two molecules. As in usual Bell inequalities scenarios, the proposed non-locality tests rely on measurements performed independently on each molecule by  observers placed far apart enough, so that no communication between them is possible during the realization of the protocol. We have shown that the proposed inequalities can be violated by a set of entangled states. An interesting point is that the inequalities derived in the present paper can also be used as entanglement witnesses, as shown, for general CHSH-tye inequalities in  \cite{WITNESS}. In this situation, one can loosen measurement conditions, since we are interested only in detecting a particular quantum correlation, and not a fundamental aspect of quantum physics. 

From the experimental point of view, motivations for this work are the recent advances in single molecule manipulation and detection with entanglement creation \cite{Hettich2002} and quantum information purposes \cite{ZOLLER, Charron2006, Yelin2006, ORIENT}.  In particular, trapped cold polar molecules are  promissing candidates for quantum information processing based on the manipulation of their rotational levels \cite{DEMILLE}. Rotational states, which are also involved in the Bell-type inequalities studied in the present paper, are relatively long lived, allowing for short quantum gate implementation times: one can hope to perform about $10^4$ gate operations before decoherence takes place. This excellent performance when compared to cold collision based quantum gates \cite{COLLI} is  due to the strength of  molecular interactions, based on dipolar forces. 

The present paper is organised as follows: in section \ref{section 1} we discuss some general properties of Bell-type inequalities that are useful in the context of the non-locality tests we propose. In section \ref{section2} we explicit the inequalities based on molecular orientation and show how it can be used for non-locality tests. We exploit its performance and study some entangled states that violate it. We then discuss some constraints involved in an eventual physical implementation of our ideas in section \ref{section3}, ending up with a concluding note in section \ref{conclusion}

\section{General Properties}\label{section 1}

We describe now some general properties of the inequalities studied in the present paper.  Our inequalities involve four observables that can be combined as follows:  

\begin{eqnarray}\label{genchsh}
&{\cal B}=&O_1(\phi_1) \otimes O_2(\phi_2)+O_1(\phi_1) \otimes O_2(\phi'_2)+\\ \nonumber
&&O_1(\phi'_1) \otimes O_2(\phi_2)-O_1(\phi'_1) \otimes O_2(\phi'_2).
\end{eqnarray}
The operators appearing in the equation above are defined as follows:  operators $O_i(\phi_i)$, $i=1,2$ are  local observables chosen by observers $1$ and $2$. The variables $\phi_i$ also depend on local properties only, and $O_i(\phi_i)=U(\phi_i)O(0)U^{\dagger}(\phi_i)$, with $U(\phi_i)$ a one-parameter group of unitary operators with periodicity $2\pi$. If the operator $O_i(\phi_i)$ is bounded, we can show that, under the assumption of local realism, (\ref{genchsh}) satisfies:

\begin{equation}\label{genuine}
|\langle {\cal B} \rangle| \leq S,
\end{equation}
where $S$ is a number, representing the maximal allowed value for a local theory to hold whenever measurements of correlations between observables are compared as in (\ref{genchsh}). The basic assumption to derive (\ref{genuine}) is that correlations can be described by probability distributions which are independent for each party ($1$ and $2$) and depend only on local parameters and that the spectrum of observables $O_i(\phi_i)$ is bounded. 
The same inequality can be derived by assuming that the average of ${\cal B}$ is taken with respect to a separable (non-entangled) states. 

For simplifying reasons, we  focus on the specific case in which  the norm of  $O_i(\phi_i)$ is the same for each subsystem and is equal to $\lambda_{max}$. In this case,   the numerical value of $S$ is $S=2\lambda_{max}^2$.

 Quantum mechanics can violate inequality (\ref{genuine}), but in order to do so,  observables $O_i$ and the state considered should be judiciously chosen. In order to check 
 whether an inequality of the type of (\ref{genuine}) allows for a  non-locality test, one should maximize the left side of (\ref{genuine}) for an arbitrary bipartite quantum state. If  the maximum obtained value is greater than $S$, all states violating (\ref{genuine}) are non-local.

Before proceeding on testing the power of inequalities of the type of (\ref{genuine}) for non-locality tests using the proposed molecular orientation-related observables, we demonstrate some simple general properties of  (\ref{genchsh}) which are independent of the observables $O_i$. Such properties are useful  since they help to simplify the numerical optimization while giving some physical insight. We show that, thanks to these properties, the number of  degrees of freedom to be considered in order to evaluate the maximal value of the violation  is decreased. We start by decomposing operators $O_i(\phi_i)$ in terms of unitary transformations. Using the group property  $U(\phi_i+\phi_j)=U(\phi_i)U(\phi_j)$ we can write:
\begin{eqnarray}\label{decomp1}
&{\cal B}=&(U_1(\phi_1)\otimes U_2(\phi_2))(O_1(0)\otimes O_2(0))+\\ \nonumber
&&O_1(0)\otimes O_2(\phi'_2-\phi_2)+ \\ \nonumber
&&O_1(\phi'_1-\phi_1) \otimes O_2(0)-\\
&&O_1(\phi'_1-\phi_1)\otimes O_2(\phi'_2-\phi_2)(U_1^{\dagger}(\phi_1)\otimes U_2^{\dagger}(\phi_2)),\nonumber
\end{eqnarray}
or, more succinctly,  
\begin{eqnarray}\label{decomp2}
&{\cal B}=&(U_1(\phi_1)\otimes U_2(\phi_2))\\
&&{\cal B}(0,0,\phi_1-\phi'_1,\phi_2-\phi'_2)(U_1^{\dagger}(\phi_1)\otimes U_2^{\dagger}(\phi_2)), \nonumber
\end{eqnarray}
with
\begin{eqnarray}\label{decomp1}
&&{\cal B}(0,0,\phi_1-\phi'_1,\phi_2-\phi'_2) \equiv O_1(0)\otimes O_2(0)+\\ \nonumber
&&O_1(0) \otimes O_2(\phi'_2-\phi_2)+ \\ \nonumber
&&O_1(\phi'_1-\phi_1) \otimes O_2(0)-\\
&&O_1(\phi'_1-\phi_1)\otimes O_2(\phi'_2-\phi_2),\nonumber
\end{eqnarray}
The interest of such decomposition is that operator ${\cal B}(0,0,\phi_1-\phi'_1,\phi_2-\phi'_2)$ depends only on two variables and possesses the same spectrum as ${\cal B}$. Thus, ${\cal B}(0,0,\phi_1-\phi'_1,\phi_2-\phi'_2)$  and   ${\cal B}$ share the same entanglement witnessing properties. Of course,  ${\cal B}$ and  ${\cal B}(0,0,\phi_1-\phi'_1,\phi_2-\phi'_2)$ do not detect the same entangled states. Nevertheless, they are connected by a local unitary transformation. 

The first question we wish to answer concerns the usefulness of orientation correlation measurement for non-locality tests. In order to answer that, it is enough to determine the norm of  ${\cal B}(0,0,\phi_1-\phi'_1,\phi_2-\phi'_2)$.  

In the next section, we will study a specific example of an operator $O_{1,2}(\phi_{1,2})$, and the physical meaning of the previous results will be made explicit. 

\section{Orientation based non-locality test}\label{section2}

\subsection{Molecular orientation and correlation}

We introduce now an essential ingredient for the inequalities studied in the present paper, which is  molecular spatial orientation. Molecular orientation is an observable, and its classical correspondent is the spatial orientation of the molecular inter-atomic axis with respect to some reference frame. By supposing that the molecules are addressed and manipulated by a linearly polarized laser field, we can define the laser's polarization axis as $z$, and take it as a reference for molecular orientation. In this case, the orientation of a given molecular state can be defined as the average value of operator $\cos{\hat \theta}$, where $\hat \theta $ is the angle relative to the $z$ axis.  Note that here, $\cos(\hat \theta)$ is taken as an operator, in perfect analogy to the position operator and related functions.  Orientation as defined above can be experimentally measured, as discussed below.

Since we are interested on probing properties related to a two party system, our system is composed by two molecules. We suppose that they behave like rigid rotors that can freely evolve. Their state depends thus on the  time $t$. The Hamiltonian describing each individual molecule's free evolution is $H_i=J_i^{2}/\hbar^{2}$, where $i=1,2$. It is expressed in units of the rotational energy. $J_i$ is the angular momentum 
operator and therefore the associated evolution operator is given by $U_i(t)= e^{-i\pi H_i t}$, where time
$t$ is written in units of the rotational period. $U_i(t)$ is therefore time-periodic with period 1. 

For each molecule of the bipartite set, the orientation at time $t$ is defined as the expectation value of 
the $\hat O_i(t)=U_2^{-1}(t)\otimes U_1^{-1}(t)\cos(\hat \theta_i)U_1(t)\otimes U_2(t)$ operator, $\moy  {\hat O_i(t)} = \bra{\psi_o}\hat O_i(t)\ket{\psi_o}$, where $\ket{\psi_o}$ is the initial state of the system. Orientation correlations between two particles are given by
the average value of 
$\moy{C(t_1,t_2)} = \moy{O_1(t_1)\otimes O_2(t_2)}$, and this quantity can be measured at different times $t_1$ and $t_2$ for each molecule. 
Operator $\cos{\hat \theta}$  is useful for entanglement detection since it ``mixes"
different values of $j$, without affecting their projection $m$. Previous works have considered correlations between different values of the projection $m$ of
a given (fixed) value of $j$ ~\cite{HD,agarwal}.

With an arbitrary accuracy, each molecule's state (subscripts  have been omitted) can  be considered to
reside in a finite dimensional Hilbert space
$\mathcal{H}$ generated by the basis set 
$\{\ket{j,m}; 0\le j\le\jmax,\abs{m}\le j\}$,  where $\ket{j,m}$ are the eigenstates of
$J^2$ and
$J_{z}$. Note that $\mathcal{H}$ has dimension $(\jmax +1)^2$. The corresponding wavefunctions are  $\braket{\theta,\varphi}{j,m}=Y_{jm}(\theta,\varphi)$, the spherical harmonics.
In the finite space $\mathcal{H}^{(\jmax)}$, the $\cos\theta$ operator is characterized by a discrete, non
degenerate spectrum of eigenvalues $\lambda_n$, with corresponding
eigenvectors $\ket{\lambda_n}$, also called {\it orientation eigenstates}. The two maximally
oriented states $\ket{+}$ and $\ket{-}$ are the two eigenstates
corresponding to the extreme eigenvalues $\pm \lambda_{max}$, where
$\lambda_{max} \equiv \text{Max}_n (\lambda_n)$. In the particular case of $\jmax = 1$ and $m=0$, maximally oriented states can be written in the basis of the angular momentum eigenstates $\ket{0,0}$ and $\ket{1,0}$ as  $\ket{+}= \sqrt{1/2}(\ket{1,0}+\ket{0,0})$ and   $\ket{-}= \sqrt{1/2}(\ket{1,0}-\ket{0,0})$. In this particular case, $\cos{\theta}=  \sqrt{1/3}\hat \sigma_x$ and we recover the results of a two-level system. In this context,  the free evolution operator $U_i(t)$ changes the orientation of a state, since orientation eigenstates  are not eigenstates of the free Hamiltonian. Time evolution creates  superpositions of  orientation eigenstates, exactly as it happens when one projects photon's polarization with polarizers. 

\subsection{The inequalities}

We can now combine all the ingredients to build the orientation based Bell-type inequalities. We start with a system composed by two molecules, and measurements independently performed in each one of them should be combined in order to tell whether the total molecular state violates or not local realism. The two molecule bipartite state $\ket{\psi_o}$ has been created at a given time $t_o$ after which it freely evolves.  We suppose that, after $t_o$, there is no interaction between molecules. 
The total molecular state is thus given by the wavefunction $\psi(\theta_1,\theta_2,\varphi_1,\varphi_2,t)\equiv \braket{\theta_1,\theta_2,\varphi_1,\varphi_2}{\psi(t)}$, where $\theta_i$ and $\varphi_i$ 
denote here the polar and azimuthal spherical coordinates in the laboratory frame. 

\begin{figure}
\includegraphics[width=0.5\textwidth]{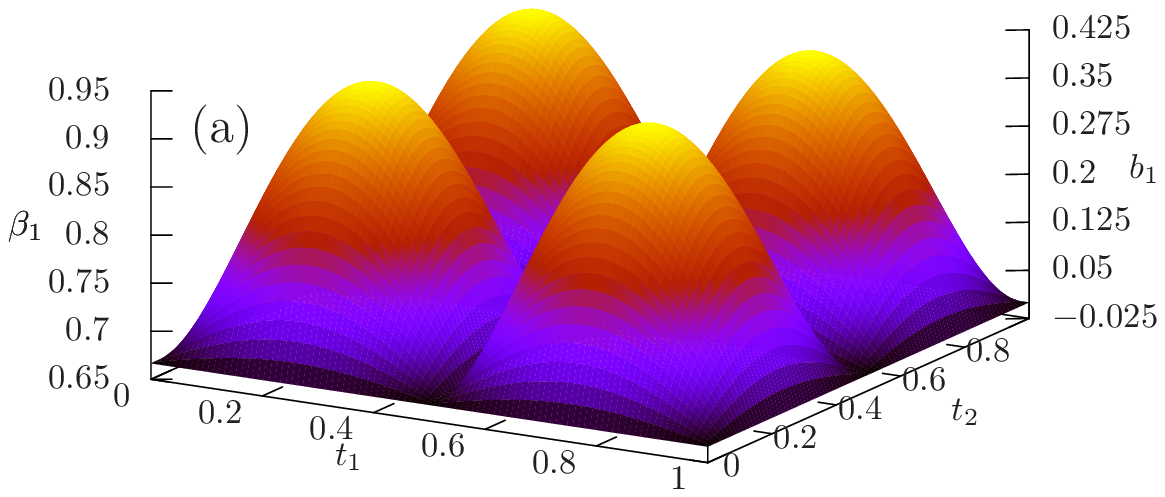}

\includegraphics[width=0.5\textwidth]{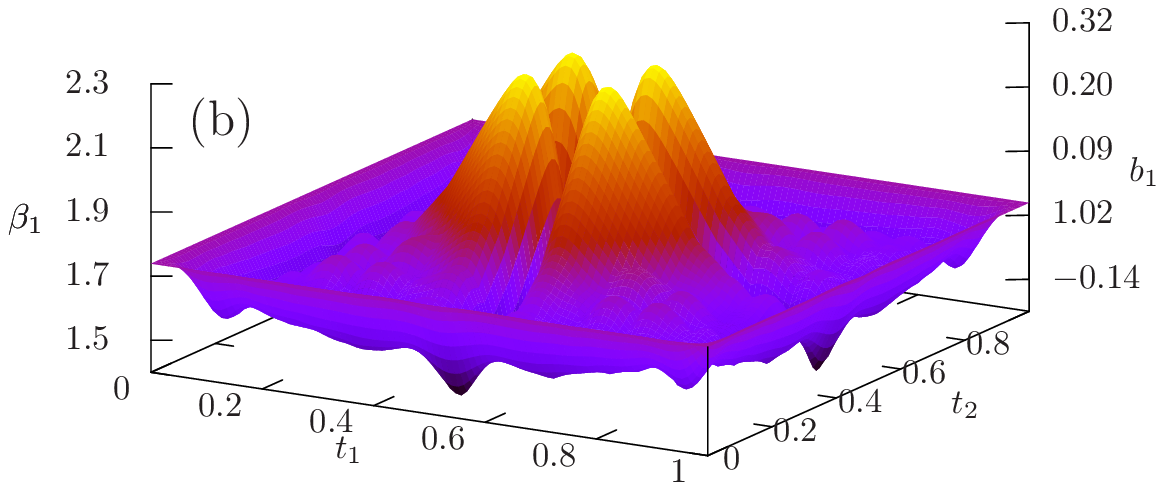}
\caption{(Color online) Maximal value of $\moy{{\cal B}_1}$ as a function of $t_1$ and $t_2$ in units of
the rotational period.
Left $z$--axis: highest eigenvalue $\beta_1$.
Right $z$--axis: relative violation $b_1$ defined by Eq.~(\ref{eq:relatviolat}).
(a): $\jmax=1, m=0$, (b): $\jmax=5, m=0$.
\label{fig1}}
\end{figure}

This state is the one whose non-local properties are to be checked. We can use it to compute the average values of $\moy  {\hat C(t_1,t_2)}$ and combine such correlations taken at different times in a way analog to  Eq.(~\ref{chsh}) and (\ref{genchsh}), defining the operator
\begin{eqnarray}
\label{cos}
{\cal B}_1(t_1,t_2,t'_1,t'_2) \equiv&& \\
&&C(t_1,t_2) + C(t_1,t'_2) + C(t'_1,t_2) - C(t'_1,t'_2).\nonumber
\end{eqnarray}
For a local theory (LT), it  obeys an  inequality similar to Eq.~(\ref{genuine}):
\beq
\label{eq:threshold}
\abs{\moy{{\cal B}_1}_{\text{LT}}} \leq 2 (\lambda_{max})^2;\quad \forall (t_i,t'_i) \in \mathbb{R}^2.  	
\eeq
Without loss of generality, we have assumed that each particle state resides in the same finite
dimensional Hilbert space $\mathcal{H}^{(\jmax)}$. Notice that, in Equation (\ref{cos}), time plays the role of polarisers in Bell inequalities based in the photonic polarisation. Other CHSH inequalities using the time evolution instead of polarizers
were studied in the literature in very different contexts: in~\cite{KAONS,GO,GISIN} they allow the detection of entanglement between products 
of decaying mesons. In~\cite{LEGGET,TEMP_DISC}, they reveal quantum properties of single particles.

\begin{figure}
\includegraphics[width=.5\textwidth]{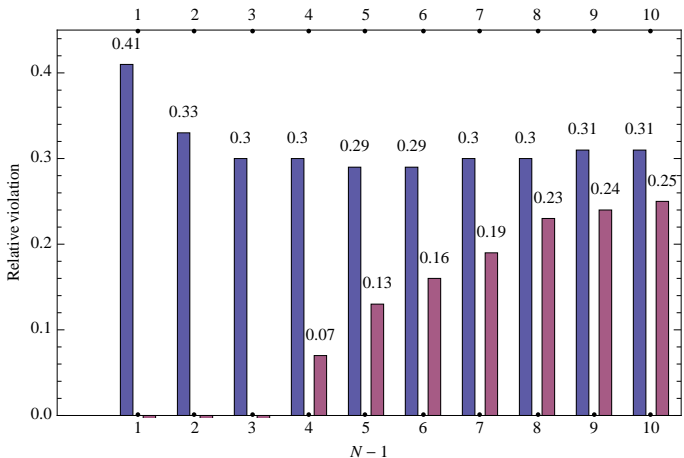}
\caption{Relative violation of the inequality (\ref{eq:threshold}) as a function of the dimension of the subspace for $m=0$. Blue: relative violation with respect to the locality threshold when only the finite dimension subspace is considered. Red: relative violation when the infinite dimensional threshold of $2$ is considered.}
\label{relative}

\end{figure}

We note that ~(\ref{eq:threshold}) is valid for all possible values of $\jmax$, 
and that it can, in particular, be extended to the limit $\jmax \rightarrow + \infty$, in which case the spectrum of $\cos{\hat \theta}$  forms a continuum. An interesting characteristic of the separability threshold ~(\ref{eq:threshold}) is its dependence on $\lambda_{max}$.

We will see in the next subsection how the general properties of CHSH-type inequalities can be used in the specific case discussed in this paper to show that the studied inequalities can be violated. 

\subsection{Reference Frame and Temporal Origin}

A straightforward application of the results derived in section \ref{section2}  consists of showing that, in order to study the spectrum of  ${\cal B}_1$, we can, without loss of generality, 
numerically diagonalize it in the specific case of $t'_1=t'_2=0$.  This result is due to the fact that both operators are  related  by a local unitary transformation. Using the notation of Section II, we have here that  $t_1=\phi'_A-\phi_A$ and $t_2=\phi'_B-\phi_B$. Another interesting application of the general results of Section II consists of showing that the inequalities discussed here allow for local realism violation even in the case where observers $A$ and $B$ have different time origins. The same happens for the spatial reference frame: violation is independent of any previous agreement between observers.  However, different temporal origins and reference frames correspond to different non-local detected states, that are related to each other by  local unitary transformations.

 We start by discussing in more details the time origin chosen by both observers. Usually, the orientation  Bell-type inequalities depend on four times of measurements, as defined in Eq. (\ref{cos}).   However, as pointed out previously, different times correspond to the application of different unitary transformations. We can thus apply the results of section \ref{section2}, identifying  the general operator $\hat O$ to the operator $\cos{\hat \theta}$.  This leads to the inequalities:
\beq
\moy { {\cal B}_1(t_1,t_2,t'_1,t'_2)}=\moy { {\cal B}_1(0,0,t'_1-t_1,t'_2-t_2)} \leq 2\lambda_{max}^2,
\eeq 
The first identity shows that states maximally violating the Bell-type inequality for $t_1,t_2,t'_1,t'_2$, defined as $\ket{s_{max}(t_1,t_2,t'_1,t'_2)}$ can be obtained by the one maximally violating it for $0,0,t'_1-t_1,t'_2-t_2$, that we will call $\ket{s_{max}(0,0,t'_1-t_1,t'_2-t_2)}$ by the application of the transformation $U(t_1) \otimes U(t_2)\ket{s_{max}(0,0,t'_1-t_1,t'_2-t_2)}$. Temporal uncertainties of $\tau_1, \tau_2$ for each one of the molecules  can always be translated as the application of the unitary operator $U(\tau_1) \otimes U(\tau_2)$, so that their only effect is to change the eigenstates of operators ${\cal B}_i$ by the same transformation. Violation can thus still be observed, and the subspace of states violating local realism are obtained by a simple unitary transformation on the original subspace.

The same type of argument can be used for uncertainties of the reference frame for each observer. All reference frames are connected by local unitary transformations describing rotations about some direction of space, and results for different references frames are connected by these same unitary transformations.  

We apply these results to simplify the investigation of possible violations of inequality (\ref{eq:threshold}). 

\subsection{Results}

 For a given value of $\jmax$, we have numerically diagonalized operator ${\cal B}_1(0,0,t_1,t_2)$, and obtained for each $(t_1,t_2)$,  
 its highest eigenvalue $\beta_1(t_1,t_2)$, which gives the maximal value of
$\moy{{\cal B}_1}$  (maximal violation of (\ref{eq:threshold})). This quantity depends on the dimensionality of the system, and we compare the amplitude of the violation when different values of  
$\jmax$ are used by defining the relative violation
\beq
\label{eq:relatviolat}
b_1(t_1,t_2) \equiv \frac{\beta_1(t_1,t_2) - 2 (\lambda_{max})^2}{2 (\lambda_{max})^2}.
\eeq
Results for different values of $\jmax$ and $m=0$ are shown in Figure \ref{fig1}. We can see that the proposed inequalities are violated for a significant region of parameters $t_1$ and $t_2$. Figure \ref{fig1} also calls one's attention because of its symmetries. The central symmetry with respect to the point $t_1=t_2=0.5$ corresponds to the time reversal symmetry. One can also easily understand the mirror symmetry about the $t_1=t_2$ line with the help of the particle  exchange symmetry of operator ${\cal B}_1$. 

We have shown in \cite{PRL} that Eq. (\ref{eq:threshold}) can be violated by a number of pure states, and numerically studied the relative violation (\ref{eq:relatviolat}) with increasing dimension. The results, shown in Figure (\ref{relative}) were obtained in a particular case, where both molecules had a vanishing angular momentum $z$ axis projection ($m=0$ for both molecules).  The effects of considering different values of $m$ will be discussed in the following. We focus here in the $m=0$ case in order to discuss the behaviour of violation of (\ref{eq:threshold}) with respect to the dimensionality of the system. We see that the maximal relative violation $b_1(t_1,t_2)$ decreases with the dimensionality of the system for low dimensions and then starts smoothly growing starting from  $\jmax=4$ (blue bars, Figure (\ref{relative})). Up to now, we have not found an asymtoptical numerical behaviour for $\jmax \rightarrow \infty$. We have shown that the maximal violation is bounded by $3$. However, it is still unknown if this value can be reached. Since the exact behaviour of the violation of inequality (\ref{eq:threshold}) in infinite dimension is still not known, one pertinent  question is   whether the proposed inequality is still violated in the limit $j_{max}\rightarrow \infty$  for physically acceptable states. We can see that it is indeed the case by calculating the violation relative to the classical bound obtained for an infinite dimensional system. In this case, the classical threshold separating local theory from non-local ones is $2$, since $\lambda_{max}=1$. Figure (\ref{relative}) shows the results of such violation (pink bars). We see that   states violating the infinite dimensional classical threshold can be  found when one considers  subspaces with dimensions higher than $N=j_{max}+1 \geq 5$ for $m=0$. In this subspace, we can see that the violation relative to the infinite dimensional threshold is still small. However, by increasing the size of the subspace and going to $j_{max}=10$, we can see that this relative violation increases, and it is not negligible when compared to the two dimensional case, for instance. Notice that the relative violation with respect to the infinite dimensional subspace and the one relative to the restricted subspace approach with increasing dimension, as one should expect. Violating the inequalities (\ref{eq:threshold}) in the infinite dimensional limit with entangled states of low dimension is a surprising result that proves that (\ref{eq:threshold}) can be violated for all possible values of $\lambda_{max}$. 

\begin{figure}

\includegraphics[width=0.5\textwidth]{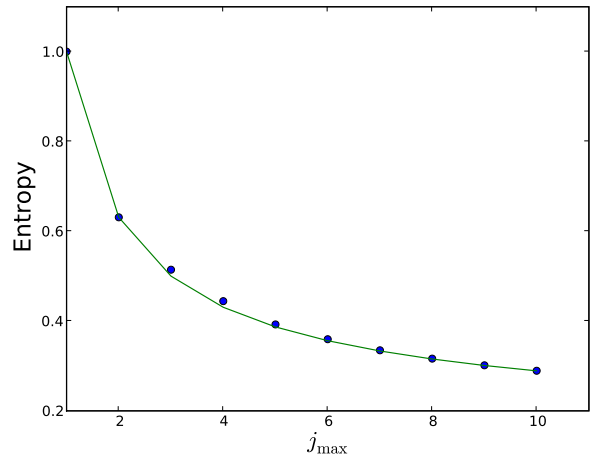}
\caption{Normalised entropy of the maximally violating state as a function of the subspace dimensionalty $\jmax$ (dots). Continuous (green) line represents $\log{2}/\log{(\jmax+1)}$.}
\label{entropy}

\end{figure}

A natural question is what states maximally violate  (\ref{eq:threshold}). Since we are dealing here with correlations in orientation, one could expect that maximally oriented  entangled states of the form $\frac{1}{\sqrt{2}}(\ket{\lambda_{max},\lambda_{max}}\pm\ket{-\lambda_{max},-\lambda_{max}})$ are those that maximally violate them. However, this is not the case, and such state, except for the case $j_{max}=1$ and $m=0$ do not violate  (\ref{eq:threshold}) at all. Nevertheless, states maximally violating  (\ref{eq:threshold}) have  most of their population in highly oriented states. For  $j_{max}=5$, states $\ket{\lambda_{max},\lambda_{max}}$ and $\ket{-\lambda_{max},-\lambda_{max}}$ carry, each one of them, $36\%$ of the population.  In addition, states maximally violating (\ref{eq:threshold}) are not maximally entangled (except in the case $j_{max}=1$, a result which is not surprising and has been observed for other Bell-type inequalities, as in \ref{ANTONIO}, for example. Using the maximally violating state, we have computed the reduced density matrix entropy $S=-Tr[\rho_i\log{\rho_i}]$, where $\rho_i$ is the reduced density matrix with respect to one of the two entangled molecules. In order to compare $S$ for different values of $\jmax$, we have normalised it with respect to $\log{(j_{max}+1)}$, which is the maximal value of the entropy in a subspace of dimension $\jmax+1$. The  reduced density matrix entropy is a measure of entanglement for pure states, as the ones we are considering here.The results are shown in Figure(\ref{entropy}), together with a plot of $\log{2}/\log{(\jmax+1)}$.  We see that the two plots almost completely match, showing that maximally violating states are very close to entangled states involving only two orthogonal states of each molecule. We have checked that the two states involved are the maximally oriented states.  

The results presented above were obtained for the specific case of $m=0$ for both molecules. Allowing $m$ to take a value different from $m=0$ does not bring qualitative changes to our results: the classical threshold will still be given by the case $m=0$, since it corresponds to the maximal value of   $\moy{{\cal B}_1}$ for a local state.   However,  the dimension of the subspace considered depends on $m$. For a given fixed $m$, it is given by $j_{max}-|m|+1$. The value of the maximal orientation is thus determined not only by the dimension  of the subspace, but also on the specific value of $m$. As a consequence, the maximal value of  $\moy{{\cal B}_1}$ is determined by the same parameters. For instance, if  the considered state is a superposition state of different $m$'s, each one of this subspaces will lead to different contributions when computing $\moy{{\cal B}_1}$. In particular, those for which $j_{max}=|m|$ give a null contribution, and the maximal possible contribution decreases as $|m|$ increases for each molecule.  Physically, this is related to the fact that states with high values of the angular momentum projection $|m|$ are less oriented than those for which $j \gg |m|$. As a conclusion, by considering different $m$'s, it is still possible to violate (\ref{eq:threshold}), even if  fulfilling  the required conditions for it becomes harder.

\subsection{Non-pure states and the effect of noise}
We now formalize the conditions a non-pure state should satisfy in order to violate Eq. (\ref{eq:threshold}). This is useful to estimate the effect of noise  in our system.  A non-pure bipartite state is described by the density matrix $\rho_{1,2}$, where the subscripts $1$ and $2$ refer to each one of the observers. The average values of  (\ref{cos}) can be obtained from $\rho_{1,2}$ by using $\moy{C(t_1,t_2)} = \moy{O_1(t_1)\otimes O_2(t_2)}= {\rm Tr}[\rho_{1,2} O_1(t_1)\otimes O_2(t_2)]$. Defining the  ${\cal B}_1$ eigenstates,  by $\{\ket{s_i}\}$, with $i$ ranging form $0$ to $(j_{max}+1)^2$ if one assumes that each particle's  subspace have the same dimension and $m=0$, it is clear that the density matrix $\rho_{1,2}$ can be expressed in this basis. By doing so,  we have that
\begin{equation}\label{rho}
{\rm Tr}[{\cal B}_1\rho_{1,2}]=\sum_i^{N^2} p_i s_i, 
\end{equation}
where $s_i$ is the eigenvalue associated to the eigenstate $\ket{s_i}$ of ${\cal B}_1$, $p_i=\bra{s_i}\rho_{1,2} \ket{s_i}$ the statistical weight of each one of such eigenstates and $N=j_{max}+1$ is the dimensionality of each molecule's subspace. This means, that for a local realistic theory, one should obey
\begin{equation}\label{rho}
\sum_{i=1}^{N^2} p_i s_i \leq 2\lambda_{max}^2. 
\end{equation}
An  example of non-pure state is given by
\begin{equation}\label{zz}
\rho_{1,2}=P_N \frac{\openone}{N^2}+(1-P_N)\ket{s_{max}}\bra{s_{max}},
\end{equation}
where $\openone$ is the $N^2 \times N^2$ identity matrix, $\ket{s_{max}}$ is the eigenstate of ${\cal B}_1$ with maximal eigenvalue $s_{max}$ and $P_N$ is the probability of the state to be in a complete mixture. This type of state has been studied  in  \cite{ZUKOWSKI} and can illustrate the presence of noise in the preparation of a state $\ket{s_{max}}$maximally violating a Bell-type inequality. They defined that the robustness of a Bell-type test with respect to noise is measured by the maximal allowed value of $P_N$ still leading to locality violation.   In the notation of Eq. (\ref{rho}), $\bra{s_{max}}\rho \ket{s_{max}}=p_{max}=P_N/N^2+(1-P_N)$ and $\bra{s_{i} \neq s_{max}}\rho \ket{s_{i} \neq s_{max}}=p_{i}=P_N/N^2$. By calculating the expectation value of ${\cal B}_1$ using (\ref{zz}) one gets that local realism should obey:
\begin{equation}\label{zz2}
P_N \sum_{i=1}^{N^2} \frac{s_i}{N^2}+(1-P_N)s_{max} \leq 2\lambda_{max}^2.
\end{equation}

\begin{figure}

\includegraphics[width=0.5\textwidth]{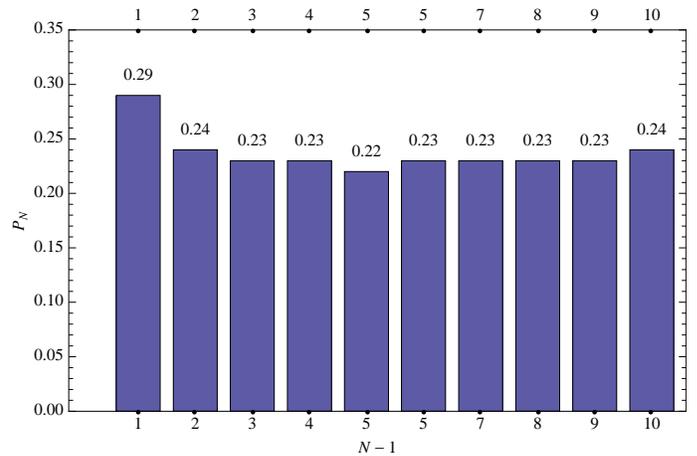}
\caption{Maximal value of $P_N$ for which our inequalities are violated as a function of $\jmax=N-1$}
\label{noise}

\end{figure}

Since ${\cal B}_1$ is traceless, $\sum_i s_i=0$, so the inequality is violated for $P_N < 1-2\lambda_{max}^2/s_{max}$. A plot of $P_N$ as a function of the considered subspace is given in Figure (\ref{noise}). Notice that, when a two-level system is considered, the known result of $P_2=1-\frac{1}{\sqrt{2}}$ is recovered. Contrary to other Bell-type inequalities for which $P_N$ was optimised \cite{ZUKOWSKI}, the maximal allowed value of noise in our system decreases with dimensionality up to $j_{max}=5$ and then starts increasing again, but doesn't change significantly in the range of dimensions that we have calculated. This result shows that the non-local properties of the maximally violating state are quite robust with respect to noise.

\subsection{Dichotomisation}\label{secdicho}

\begin{figure}

\includegraphics[width=0.5\textwidth]{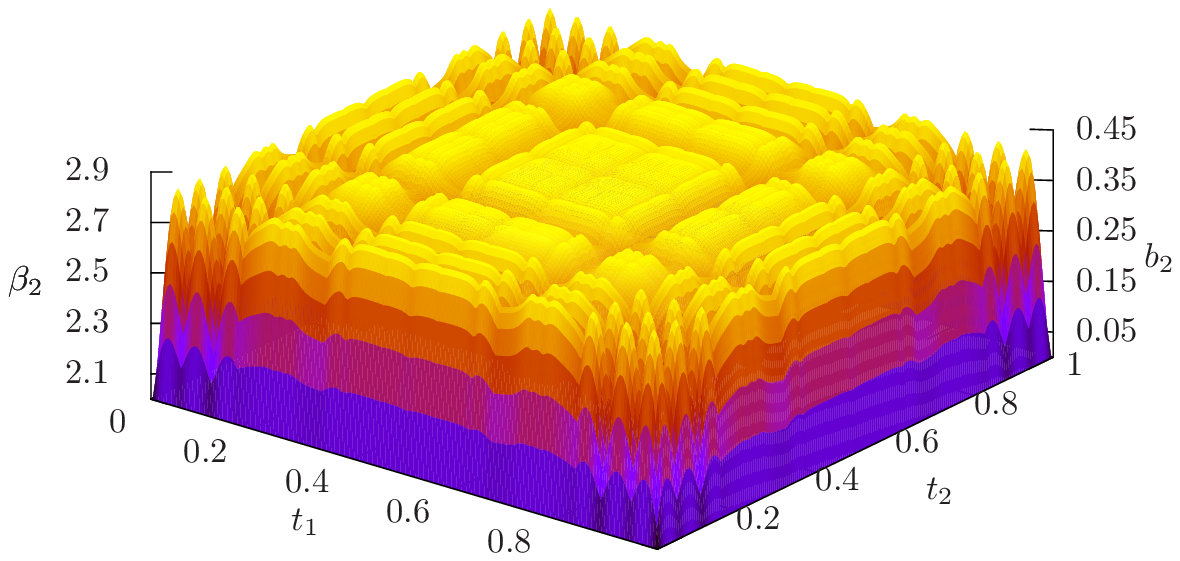}
\caption{(Color online) Same as Fig.~\ref{fig1}.b, but for $\moy{{\cal B}_2}$}
\label{dicho}

\end{figure}

We have shown that  the operator defined by Eq.~(\ref{cos})  allows not only for the realization of Bell-type tests in finite angular momentum subspaces, but also when the size of the subspace is not a priori known. This proves that, even in the case where we consider an infinite dimensional space, our inequalities can be violated for some states. However, as shown previously, the contrast of the maximal violation depends on the value of $j_{max}$. With dichotomizing procedure,  a high dimensional system is transformed into an effective two level one by splitting in two the set of measurement outcomes, and the contrast is kept constant, equal to the one for a two-level system, irrespectively of $j_{max}$. In our case, we can dichotomise  as follows: 
states $\ket{\varphi}$, for which $\moy{\cos\theta}_{\varphi}>0$, are said to be positively oriented, while
those for which $\moy{\cos\theta}_{\varphi}\le 0$ are considered as negatively oriented.
We  define the associated projectors: 
$\Pi_{\pm}=\sum_{\lambda_{\pm}}\ket{\lambda_{\pm}}\bra{\lambda_{\pm}}$ which project states
in the subspace of positive (negative) orientation.
$\lambda_{\pm}$ are the positive (negative) $\cos{\theta}$ eigenvalues in the given
finite dimensional subspace and $\ket{\lambda_{\pm}}$ are the corresponding eigenstates. The
measured observable for each molecule $i$ is then $\Pi_i=\Pi_+-\Pi_-$. 
For a given single-molecule state
$\ket{\varphi}=\sum_{\lambda_+}c_{\lambda_+}\ket{\lambda_+}+\sum_{\lambda_-}c_{\lambda_-}\ket{\lambda_-}$,
$\langle \Pi \rangle_{\varphi}$ can take any value in the interval $[+1,-1]$. The total two molecule observable is  $\Pi=\Pi_1\otimes \Pi_2$. 
We refer, as before, to two molecule correlation measurements realized at two different 
times, using $\Pi(t_1,t_2)=\Pi_1(t_1)\otimes\Pi_2(t_2)$ where $\Pi_i(t_i)=U_i^{-1}(t_i)\Pi_iU_i(t_i)$.
In analogy with Eq.~\ref{cos}, we now define the operator
\begin{equation}
\label{eq:B2}
{\cal B}_2=\Pi(t_1,t_2)+ \Pi(t_1,t'_2)+\Pi(t'_1,t_2)- \Pi(t'_1,t'_2).
\end{equation} 
Since $\Pi(t_i,t_j)^2=1$, one can show that the highest value $\moy {{\cal B}_2}$ can reach is given by the Cirel'son bound $2\sqrt{2}$~\cite{CIRELSON}. Also, for a local theory, we have
\beq
\label{dico} 
\abs{\moy{{\cal B}_2}_{\text{LT}}} \leq 2;\quad \forall (t_1,t_2) \in \mathbb{R}^{2}. 
\eeq
As in the case of ${\cal B}_1$, we can define the relative violation $b_2(t_1,t_2)$ as a function of the maximal eigenvalue $\beta_2(t_1,t_2)$ of ${\cal B}_2(t_1,t_2)$ (see Equation (\ref{eq:relatviolat})).  The interest of dichotomisation is that the maximal value of  the relative violation is always given by $b_2(t_1,t_2)=\sqrt{2}-1$,  independently of the dimension  of the subspace. Fig.~\ref{dicho} shows the maximum value $\beta_2(t_1,t_2)$ of $\moy{{\cal B}_2(t_1,t_2)}$ 
as a function of $t_1$ and $t_2$ for $\jmax=5$, while for  $\jmax=1$ we obtain trivially 
the same result as with $b_1(t_1,t_2)$ (Fig.~\ref{fig1}.a right z--axis). We see in Fig. \ref{dicho} that the dichotomisation procedure not only keeps the contrast of the maximal violation constant, but also allows for high violation for a wider range of values of $t_1$ and $t_2$.

\section{Discussion on possible experimental implementations}\label{section3}
Orientation entanglement between two molecules can be created, for instance, with the help of the dipolar interaction \cite{Charron2006}. A possible experimental scenario consists of two trapped molecules that can be submitted to spatial displacements, as it is currently done with atoms \cite{BLOCH}. By putting molecules close enough to each other, and with the help of laser manipulation, one can tailor rotational entangled states. In order to realise Bell-type tests, molecules would have to be separated and measurements performed independently in each one of them. The proposed scenario is  realistic and there has been rapid progress on cooling, trapping and manipulating polar molecules, especially those formed with two alcali atoms. We believe that such manipulations are going to be shown in laboratories in a near future.  With respect to the tests we wish to perform in the present paper,  there are some crucial points to analyse for the  experimental implementation of the proposed Bell-type inequality test. One of them  concerns trapping conditions. Traps should allow molecules to make spatial displacements so that they are sufficiently far apart and so that signaling of one observer's results to the other is not possible during at least one molecular rotational period. This condition is attained for an inter-molecular distance $L > cT$, where $c$ is the speed of light and $T$ is the molecular rotational period. To put some numbers on it, the molecular period is usually of the order of $10^{-12}$s, giving a lower bound to $L \approx 10^{-6}$m. Even if at this distance dipolar interaction between molecules is not completely negligible, measurements on both molecules are performed during the same rotational period, which is much faster than the characteristic interaction time related to dipolar coupling at this distance (see \cite{Charron2006}, for example, for a detailed discussion on the relevant time scales). Optical traps seem  promising candidates to trap and to displace molecules. Optical tweezers \cite{GRANGIER1}, optical lattices and optical conveyor belts \cite{ARNO1} are some examples of dipole force based optical traps allowing for atomic trapping and coherence preserving displacements \cite{GRANGIER2, ARNO1, BLOCH}. In the case of optical tweezers, highly focused beams are used to trap individual atoms \cite{GRANGIER1}, and we can, in principle,  load two atoms, each one of them in an individual trap \cite{GRANGIER3}. In optical conveyor belts, a stationary field made of two counter propagating beams is used to trap atoms, that can be displaced over distances of $\approx 10$nm by changing the relative detuning between the laser beams \cite{ARNO1}. Going from atoms to molecules in this type of experimental system could in principle be done by the usual photo association techniques, already demonstrated in the context of optical lattices \cite{MOLECULES}, where diatomic molecules of two alcali atoms of the same specie have been produced. Once  molecules are created and trapped, they interact with each other by dipole interaction and a non-local state may be created, as discussed in the previous section. Molecules are thus separated until dipole interaction is negligible, and measurements can be performed. One natural question concerns the effect of the trapping potential itself to the orientation of the molecular state. Optical traps are based on an applied non resonant electric field that interacts with rotational levels. We suppose that molecules in our setup are created in their ground electronic level. The optical trap is created by coupling non-resonantly this electronic level to the first excited one. The detuning between laser and the electronic transition is $\delta$. In usual experiments on optical tweezers, for instance,, $\delta$ can vary vary in the range  $\approx 10-10^4$GHz, providing photon emission rates from the excited electronic state in the range of $0.1-100$ MHz. These figures are of the same order of magnitude of coherence lifetime for rotational levels, while the range of variation of $\delta$ is also of the same order of magnitude of the frequency difference between neighboring rotational levels. In order to estimate various effects of trapping lasers, we will put ourselves in a close to realistic configuration in which the rotational frequency is $\approx 10$GHz and $\delta \approx 10^4$GHz. Each rotational level has a different energy, making the effective detuning $j$ dependent. This corresponds to adding a factor $\delta_j=B j(j+1)/\hbar$ to the electronic detuning $\delta$, taken with respect to the rotational ground state. Because each rotational level has a detuning which depends on the value of $j$,  there will be a phase difference between each rotational level due to the non resonant coupling to light. In the case of a harmonic trapping potential, this phase is given by $e^{-i 2\Omega^2/(\delta+\delta_j) t}$ if only one electronic transition is considered. This corresponds to a local action on each molecule that does not play a significant role on non-locality tests, since it maps one state to another one with the same degree of entanglement. As shown before, by properly choosing the time where measurements are performed, such effect can be compensated. This means that  the parameters for which maximal violation occurs may be modified.  As it can be shown, if a given state violates the inequalities of the type of (\ref{eq:threshold}) and (\ref{dico}), a collection  of states  connected to it by local unitary transformations will also violate the same inequalities, even if for different parameters. Also, it is important to notice that such dephasing effects can be completely neglected depending on the precise circumstances under which the experiment is performed. This happens because dephasing occurs on a time scale much longer than the rotational period, so that if measurements are performed in this interval, it will not affect the expected results.

Finally, we address the question of orientation measurement. A possibility is to use Coulomb explosion, a destructive techniques already employed for molecular orientation measurements \cite{SAKAI}. After a quasi-instantaneous  (when compared to the molecular rotational period), laser induced dissociative multi-ionisation, the molecular fragments are recorded at different directions of space, in the case of the orientation Bell-type inequalities, or at different hemispheres, in the case of the dichotomized inequalities. This technique is still experimentally challenging since it demands highly efficient single atom detectors. Usually this technique is employed for molecular ensembles, and the detection efficiency is less determinant than in our case. Another possibility is to detect the orientation of single molecules optically, as realized in \cite{NIST}. Fluorescence intensity when a single molecule is excited by a linearly polarized laser beam that can change polarization in time, can reveal molecular orientation since the scalar product between the field polarization and the molecular dipole depends on its orientation. The phase of the fluorescence intensity with respect to the exciting laser beam when its polarization is turned in time, depends on the molecular orientation, as observed in \cite{NIST}.

 \section{Conclusion}\label{conclusion}
 
We have extensively discussed some important properties of a recently proposed Bell-type inequality based on the measurement of a continuous, bounded, observable, which is molecular orientation. We have discussed some important properties of it, as its symmetries and its non-dependency on a specific choice of common temporal origin or reference frame. The role of noise in our system, that would transform pure states into statistical mixtures was also analysed, as well as how violation depends on the amount of noise. We have also discussed some important conditions that an eventual experimental set-up should satisfy and the influence of physical parameters as light forces and residual interaction between molecules on the proposed measurements.  It seems, from our analysis, that once the necessary degree of advancement is attained by experimental set-ups, our proposal is realistic and should not present major difficulties.  Our results open the perspective 
of entanglement detection and non-locality tests for high angular momentum systems in atomic and molecular physics.

One interesting aspect of the derived inequalities is the fact that they assume the simple form of CHSH inequalities and deal with continuous variables at the same time. This happens because  the observable that is measured for each particle has a bounded spectrum, naturally limiting the locality threshold and the norm of the Bell operator. Notice that, in spite of having discussed here the specific case of molecular orientation, the same type of inequalities could be derived for other continuous bounded observables, easily measurable in other physical contexts. 

\begin{acknowledgments}

This work was partially supported by  ANR  (Agence Nationale de la Recherche), project ImageFemto number ANR-07-BLAN-0162-02 \end{acknowledgments}

\end{document}